# Orbital Hall Effect Enables Field-Free Magnetization Reversal in Ferrimagnets without Additional Conversion Layer

Zelalem Abebe Bekele[1], Kun Lei[1,2], Xiukai Lan[1], Xiangyu Liu[1,2], Hui Wen[1,2], Weihao Li[1,2], Yongcheng Deng[1], Wenkai Zhu[1,2], Kaiming Cai[3], Lishu Zhang[4,*] and Kaiyou Wang[1,2,*]

[1] *State Key Laboratory of Semiconductor Physics and Chip Technologies, Institute of Semiconductors, Chinese Academy of Sciences, Beijing 100083, China*
[2] *Center of Materials Science and Optoelectronics Engineering, University of Chinese Academy of Sciences, Beijing 100049, China*
[3] *School of Physics, Huazhong University of Science and Technology, Wuhan 430074, China*
[4] *School of Materials Science and Engineering, Shandong University, Jinan 17923, China*

*Email: lishu.zhang@sdu.edu.cn, kywang@semi.ac.cn

## ABSTRACT

The spin Hall effect provides a well-established route for electrical magnetization control, while the orbital Hall effect offers a powerful yet less explored source of angular momentum. Achieving field-free deterministic switching in straightforward orbital-torque architectures remains challenging. Here, we demonstrate orbital-Hall-current-driven switching in a Mo/CoGd bilayer without the need for a separate orbital-to-spin conversion layer across a wide temperature range. In this simplified geometry, Mo serves as both an orbital and spin current source. However, the spin contribution is insufficient due to weak spin-orbit coupling, which is consistent with first-principles calculations predicting a large orbital Hall conductivity. The adjacent ferrimagnetic CoGd layer provides both orbital-to-spin conversion and the perpendicular switching medium. Planar Hall and current-induced loop-shift measurements reveal a substantial unconventional *z*-polarized damping-like torque originating from interfacial symmetry breaking. Increasing the Mo thickness from 0.2 to 2 nm increases torque efficiency by approximately 31% (*y*-polarized) and 71% (*z*-polarized) components. This enhancement enables field-free deterministic switching with a critical current density down to $2.51 \times 10^6$ A cm$^{-2}$. Our results establish Mo/CoGd bilayers as a compact platform for orbital-current switching and point toward low-power orbitronic memory devices.

Spin-orbit torque (SOT) provides an efficient electrical means of controlling magnetic order in nanoscale devices and has become a central mechanism in non-volatile spintronic memories.[1-4] Conventionally, SOT switching is driven mainly by spin currents generated through the bulk spin Hall effect (SHE) in heavy metals or via interfacial Rashba-Edelstein effects.[2-5] Although these mechanisms have enabled current-induced switching of perpendicular magnets, deterministic field-free operation generally requires either an external assist field, strong spin-orbit coupling (SOC) materials, or additional symmetry-breaking design.[2-12]

The orbital Hall effect (OHE) offers a complementary strategy for angular momentum transfer generation beyond conventional SHE materials.[13-18] Since pronounced orbital currents can arise even in weak SOC metals, orbital-based transport substantially extends the range of candidate materials for current-induced magnetization control.[15-25] Direct magneto-optical observation of orbital accumulation in Ti has further established orbital angular momentum (OAM) as an active transport degree of freedom in solids.[18] A central challenge, however, is that OAM must be converted into spin angular momentum (SAM) before it can exert a torque on a magnetic order parameter, and its efficiency depends strongly on the specific conversion layer.[13,19,20,26] Thus, orbital-torque efficiency ($\xi_{OT} = \theta_{OHE} \cdot \eta_{L-S}$) is governed not only by orbital Hall response ($\theta_{OHE} = (2e/\hbar)\,\sigma_{OH}/\sigma_{xx}$) of the source layer, but also by orbital-to-spin conversion capability ($\eta_{L-S}$) of the adjacent magnetic layer, interface, or inserted conversion layer.[19,20,27]

Understanding how orbital currents propagate and convert into spin currents is essential, yet it remains challenging due to the difficulty of disentangling orbital and spin dynamics in thin films. This conversion process fundamentally governs the design of orbital-torque devices.[16-20,27,28] Previous studies have enhanced orbital-to-spin conversion efficiency through strong SOC interface engineering, employing strategies such as Pt insertion, rare-earth layers, oxides, or magnetic stacks.[17-19,24,27,28] These approaches have enabled efficient orbital Hall torques and, more recently, the switching of perpendicular magnetic layers using orbital Hall materials like Zr.[20,23] Recent work on Ti/ferrimagnet bilayers has further revealed that rare-earth transition-metal (RE-TM) ferrimagnets can enhance orbital-torque switching by boosting the SOC-related conversion capability.[24,28] These developments underscore the promise of orbital-current-driven switching, achievable by adjusting the magnetization from RE-dominated to TM-dominated (and vice versa) across a wide temperature spectrum.[19,20,24,27,28] Nevertheless, a critical question remains unresolved: can the direct injection of orbital current achieve field-free deterministic switching in a straightforward OHE/ferrimagnetic bilayer configuration over a broad temperature range, where the ferrimagnetic layer concurrently serves as both the switching and the orbital-to-spin conversion medium?

Mo/CoGd provides a suitable platform to address this question. Mo is a promising orbital-current source because 4*d* transition metals, including Mo, are predicted to host large orbital Hall conductivity ($\sigma_{OH}$ = 3.83 ($10^3$ $\hbar$/e)($\Omega$ cm)$^{-1}$) compared to their spin Hall conductivity ($\sigma_{SH}$ = -0.25 ($10^3$ $\hbar$/e)($\Omega$ cm)$^{-1}$), with relatively low-resistance current transport.[14,21,25,29,30] Antiferromagnetically coupled ferrimagnetic CoGd is an attractive magnetic counterpart because it supports perpendicular magnetic anisotropy (PMA), tunable magnetic compensation characteristics, and efficient current-induced dynamics.[26,27,31-37] Notably, Gd sites in RE-TM ferrimagnets facilitate highly efficient local orbital-to-spin conversion. Leveraging this unique characteristic, antiferromagnetically coupled ferrimagnetic CoGd can simultaneously act as the perpendicular ferrimagnetic switching layer and an intrinsic medium for orbital-to-spin conversion within the device system.[24,26,27,28]

Deterministic field-free switching of a perpendicular magnet necessitates a torque component that breaks the symmetry between the two out-of-plane magnetic states.[7-12] In the Mo/CoGd bilayer, interfacial inversion-symmetry breaking is accompanied by the chemically inequivalent and antiferromagnetically coupled Co-3$d$ and Gd-4$f$ sublattices. From a symmetry perspective, this reduced magnetic symmetry removes the operation that would otherwise map $+m_z$ into $-m_z$ under a fixed in-plane current. Consequently, current-induced orbital/spin torque components can act as an intrinsic symmetry-breaking bias, selecting a unique switching polarity and enabling field-free deterministic switching in the Mo/CoGd system.

Here, we demonstrate orbital-current-induced field-free deterministic switching in perpendicular Mo/$Co_{71}Gd_{29}$ ferrimagnetic bilayers without a separate orbital-to-spin conversion layer across a wide temperature range. Through systematic comparisons with control samples, together with first-principles calculations and a phenomenological thickness-dependent model (Supplemental Materials S1), we identify the roles of strong orbital-current injection and interfacial symmetry breaking. Enhanced planar Hall effect (PHE) and current-loop-shift measurements reveal a sizable, unconventional $z$-polarized damping-like torque associated with interfacial symmetry breaking. Together with first-principles calculations showing an orbital-dominated Hall response in Mo, these results demonstrate a compact route to orbital-current-induced, field-free, deterministic switching in ferrimagnetic bilayers.

Antiferromagnetically coupled Co and Gd sublattice moments exhibit different temperature dependences, leading to a compensation temperature $\left(\mathrm{T}_{Comp}\right)$ in CoGd ferrimagnet.[6] The compensation temperature is sensitive to the composition of CoGd alloy.[27,37] Figure 1 shows the measured anomalous Hall resistance ($R_H$) hysteresis loops for the Mo/CoGd sample under out-of-plane magnetic field $\boldsymbol{H_z}$ at temperatures below $\mathrm{T}_{Comp}$ (100 K, Gd-dominated sublattice) (Fig. 1a), near $\mathrm{T}_{Comp}$ (137 K) (Fig. 1b), and above (300 K, Co-dominated sublattice) (Fig. 1c). At 137 K, the Co and Gd moments fully compensate ($|\boldsymbol{m}_{Co}| = |\boldsymbol{m}_{Gd}|$), resulting in zero net magnetic moments and therefore zero $R_H$. Across the $\mathrm{T}_{Comp}$, the coercivity diverges and the loop polarity reverses (Fig. 1d), indicating a switch in magnetic dominance between the Gd and Co sublattices. The magnitude of $R_H$ as a function of temperature (Fig. 1d) reaches its minimum value at the $\mathrm{T}_{Comp}$.

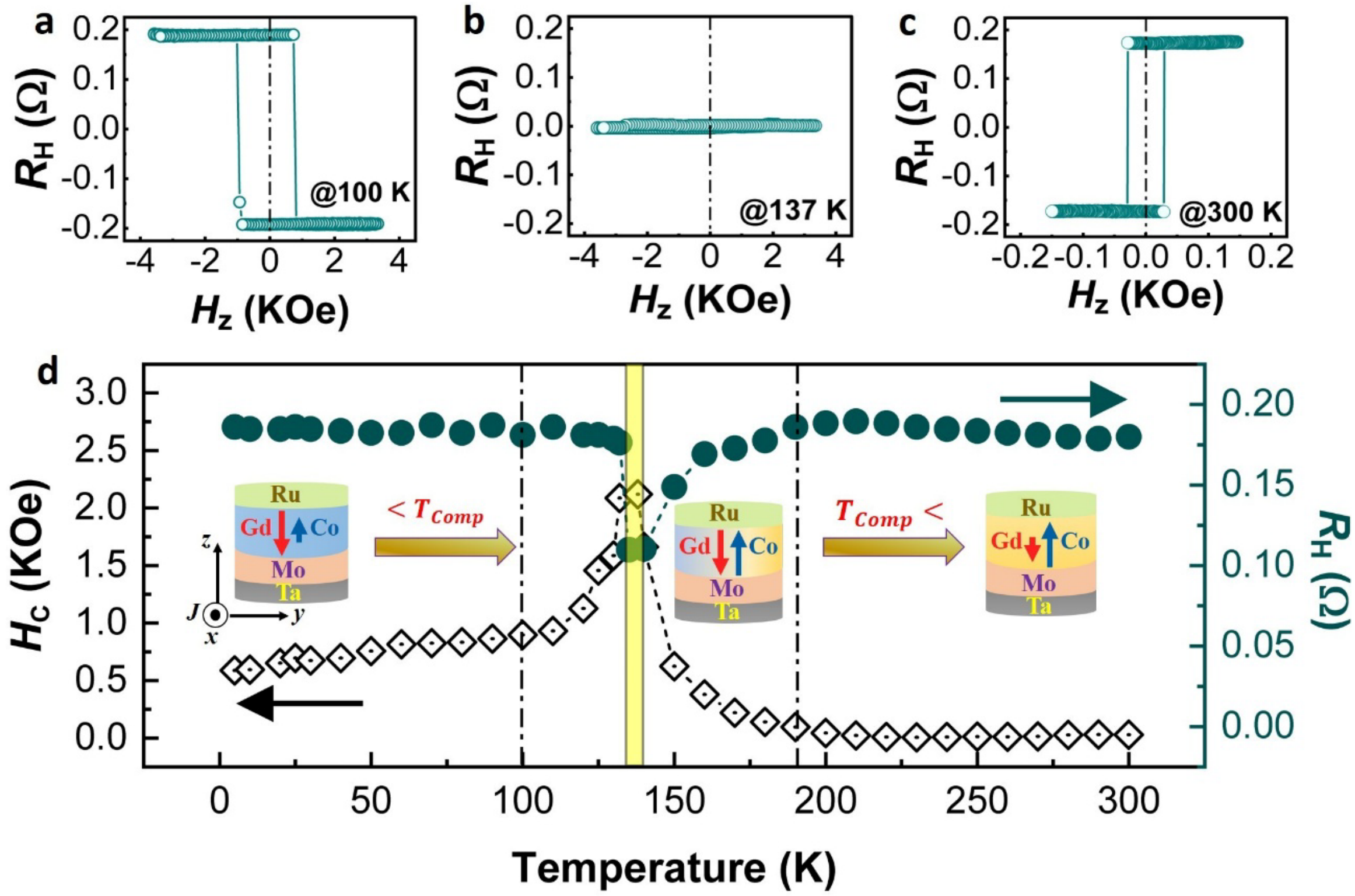

**FIG. 1. Magnetotransport properties with different temperatures for the Hall bar devices of Mo(1)/CoGd(5) stacks with a small source current of 50 μA along the *x* direction.** (**a-c**) The anomalous Hall resistance $R_H$ as a function of an out-of-plane magnetic field $\boldsymbol{H}_z$, measured for temperatures below (100 K) (**a**), around (137 K) (**b**), and above (300 K) (**c**) the compensation temperature. (**d**) Temperature dependence of coercivity field (open black circle) and the extracted anomalous Hall resistance $R_H$ (solid dark cyan circle) as a function of temperature. The inset schematic arrow illustrates the stoichiometry-weighted (spin and orbital) magnetic moments of Gd and Co, along with a temperature-controlled reversal of the net magnetization.

Figure 2a shows a schematic illustration of the spin and orbital currents generated by the Mo layer and then injected into the neighboring CoGd, where Gd's strong SOC converts the incoming $\boldsymbol{J}_{OHE}$ into a spin current. Field-free current-induced opposite-polarity deterministic switching is achieved at 20 K (clockwise) and 300 K (counterclockwise) (Fig. 2b and 2c). Notably, the critical switching current density at 20 K is higher than that at 300 K. At 20 K and 300 K, the estimated Joule heating-induced device temperature $(T_d)$ rise from the large switching current density is about 90 K and 45 K, respectively (Supplemental Material S2). Compared with the switching at 300 K, the reduced current-driven normalized $R_H = R_H^{Ip}/R_H^{Hz}$ (60%) at 20 K indicates incomplete switching, which can be attributed to the pinning effect or to more pronounced Joule heating at lower temperatures.[38,39] The opposite-polarity switching across $\mathrm{T}_{Comp}$ arises from the net magnetization dominated by antiparallel Gd and Co spins. Likewise, measurements under an in-plane field further validate that the current-driven deterministic switching polarity is reversed on either side of $T_{Comp}$ (Supplemental Material S3), consistent with a previous report.[40] The pulsed-width current-induced repeatable switching highlights the reproducibility of all-electrical switching in Mo/CoGd (Supplemental Material S4).

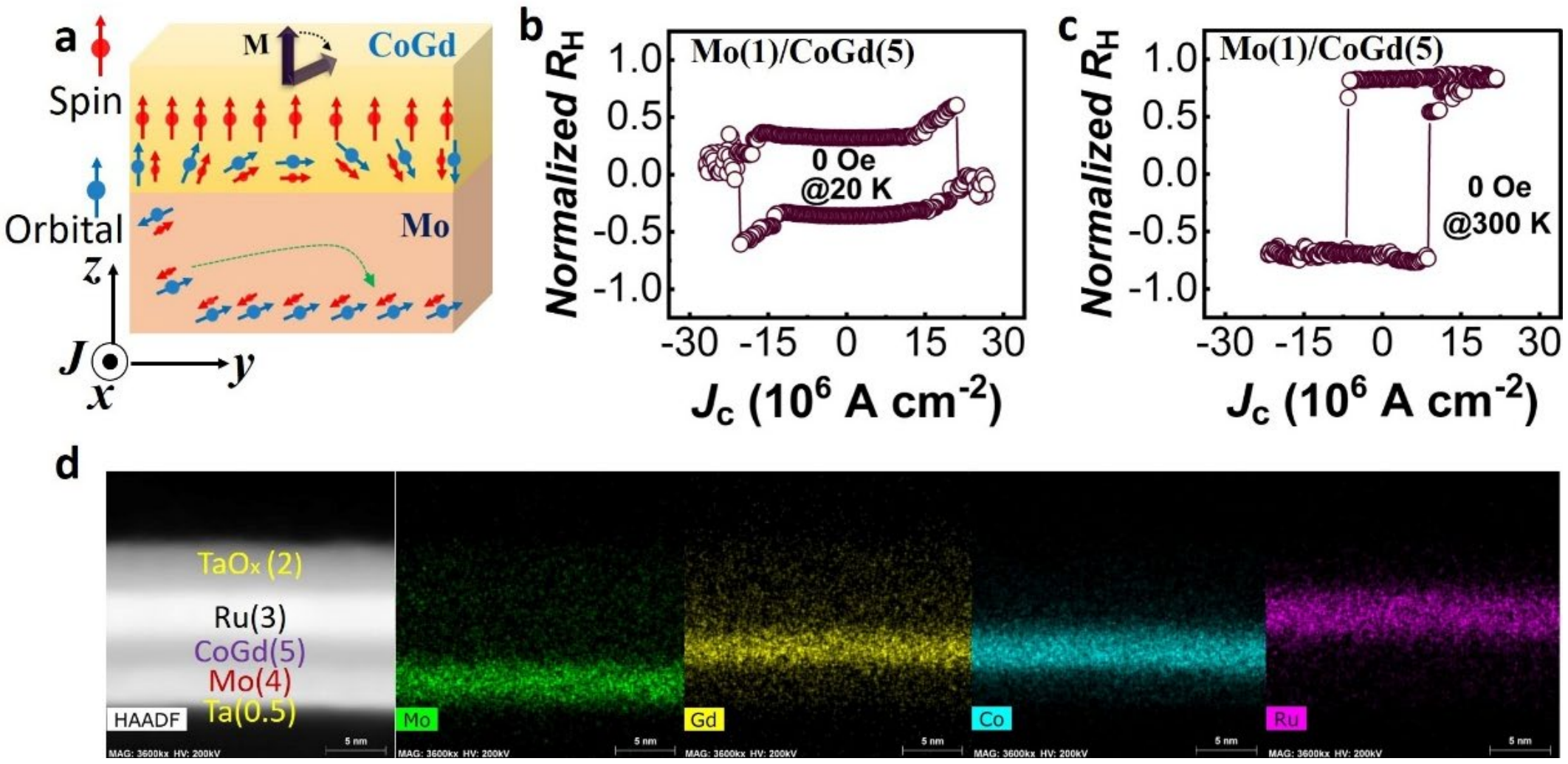

**FIG. 2.** (**a**) The schematic illustration of the Mo/CoGd stack structure, where the bottom Mo layer induces relatively strong $\boldsymbol{J}_{OHE}$ as compared to their spin current ($\boldsymbol{J}_{SHE}$) and the CoGd layer converts the incoming $\boldsymbol{J}_{OHE}$ into $\boldsymbol{J}_{SHE}$. The blue and red arrows represent the orbital and spin angular momentum, respectively. A large black-and-purple arrow illustrates the local CoGd magnetic moment and charge current, respectively. (**b-c**) The changes in the normalized $R_H$ as a function of switching current density $J_c$ for the Hall bar device in the absence of a magnetic field ($H_x = 0$ Oe) for 20 K (**b**) and 300 K (**c**) for Mo(1)/CoGd(5) bilayer devices. (**d**) The high-resolution scanning transmission electron microscope (HRSTEM) and the energy dispersive spectroscopy (EDS) mapping of the Mo(4)/CoGd(5) bilayer sample.

To verify whether the field-free deterministic switching comes from the composition gradient, high-resolution scanning transmission electron microscopy (HR-STEM) with energy-dispersive X-ray spectroscopy (EDS) was performed. Sharp interfaces and homogeneous element distribution within their own layers indicate negligible intermixing (Fig. 2d and Supplemental Material S5), ruling out the composition gradient induced by the symmetry-breaking.[41]

It is well known that both in-plane symmetry breaking and competing spin polarization can induce an out-of-plane effective $H_z$ field,[7-12,39] resulting in field-free current-induced deterministic switching. In CoGd alloys, the antiparallel spin orientations of the Co and Gd sublattices generate a spin polarization that could potentially enable field-free deterministic switching through a competing spin current mechanism.[35] To distinguish the role of this intrinsic competing contribution, we fabricated Pt(1.8)/CoGd(5)/Pt(1.5) and Pt(4)/CoGd(5)/Ru(3) control samples (Supplemental Materials S6 and S7). Pt-based control samples with robust PMA cannot achieve field-free deterministic switching, ruling out competing spin effects as the underlying switching mechanism.

To further verify the origin of the orbital current and rule out any contribution from the top Ru layer, a Mo(2)/CoGd(5)/$AlO_x$(3) was fabricated without the Ru cap. This Ru-free sample retains robust PMA and exhibits field-free deterministic switching (Supplemental Materials S8), confirming that the orbital current in our Mo/CoGd system predominantly originates from the Mo layer.

To better understand the microscopic origin of the transverse angular-momentum response of Mo in our Mo/CoGd system, we performed first-principles calculations of the orbital Hall conductivity (OHC) and spin Hall conductivity (SHC), along with their momentum-resolved distributions across the two-dimensional Brillouin zone of the Mo slab. The calculated OHC consistently exceeds the SHC over a broad energy window centered on the true Fermi level. At $E_F - E_F^{\mathrm{true}} = 0$, both conductivities are positive, with the OHC approximately twice the magnitude as compared to that of the SHC, indicating a predominantly orbital nature of the Hall response in the Mo slab (Fig. 3a). As the Fermi level shifts toward higher energies, both conductivities gradually decrease and undergo a sign reversal near $E_F - E_F^{\mathrm{true}} \sim 1.0$ eV, yet the OHC remains systematically stronger than the SHC throughout the investigated energy window, including on the electron-doped side, where the negative excursion of the OHC is substantially enhanced. Complementing the energy dependence, we further analyzed the $k$-resolved orbital texture and OHC distributions of $J_y^{L_z}(\mathbf{k})$ on the Fermi surface (Fig. 3b and Supplemental Material S9). The $k$-resolved orbital texture maps reveal pronounced hot spots near the Brillouin-zone boundary, particularly around the $X$ points, while contributions near the $\Gamma$ point remain comparatively weak. Notably, compared with the spin Hall response, the orbital Hall response exhibits not only a larger peak intensity but also extends over a significantly broader region of momentum space. Collectively, these first-principles calculation results confirm that the orbital channel is activated much more strongly than the spin channel in Mo, providing a robust microscopic foundation that strongly supports the experimental findings.

Crucially, the critical switching current density ($J_c$) decreases significantly as the Mo layer thickness increases up to 4 nm in the Mo($y$)/CoGd(5) bilayer, reaching a remarkably low value of $2.51 \times 10^6$ A cm$^{-2}$ at 300 K for Mo(4)/CoGd(5) (Fig. 3c). This represents an order-of-magnitude reduction compared to systems that rely on explicit SOC conversion layers, such as Cr/Pt/CoFeB ($7.50 \times 10^7$ A cm$^{-2}$),[19] and $CuO_x$/Pt/TmIG ($1.70 \times 10^7$ A cm$^{-2}$).[42] The reduction in switching $J_c$ is attributed to the enhanced damping-like torque and decrease in coercivity field in our system.[43]

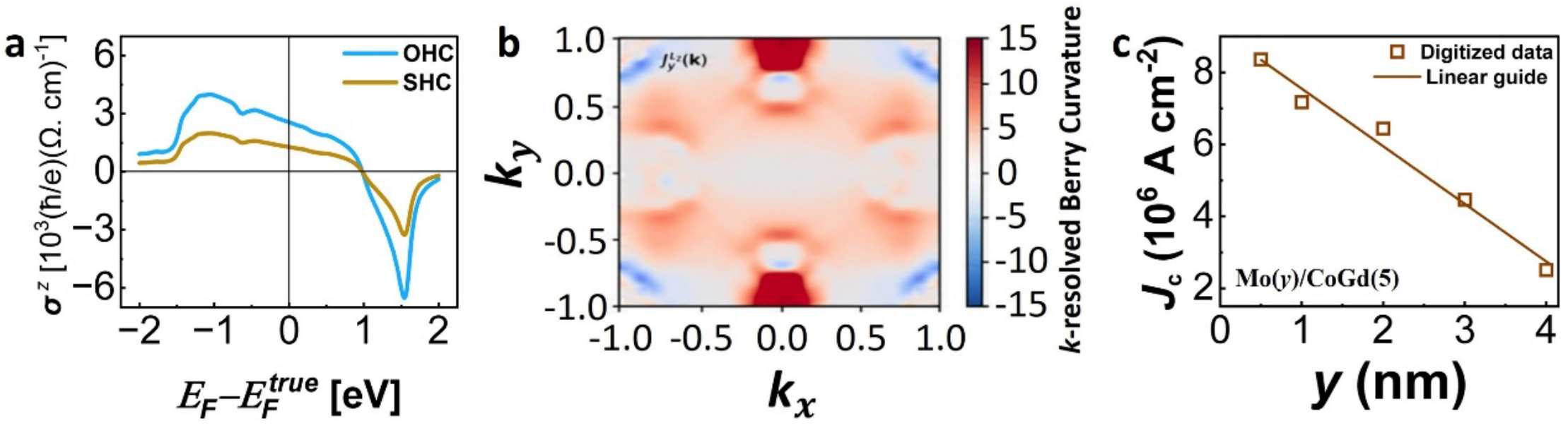

**FIG. 3.** (**a**) The first-principle calculations of OHC and SHC as a function of energy relative to the true Fermi level. Normalized orbital-source factor $F_L$ and PHE-derived interfacial factor $B_{\text{int}}$. (**b**) $k$-resolved intrinsic OHC distributions of $J_y^{L_z}(\mathbf{k})$ on the Fermi surface of the Mo slab. The dominant hot spots are concentrated near the zone-boundary $X$ points, while the response around Γ is relatively weak. (**c**) Critical switching current density $J_c$ for a series of Mo(*y*)/CoGd(5) samples with *y* varying from 0.5 to 4 nm, where the linear guiding fit.

The first- and second-harmonic Hall signals were measured to determine the damping-like ($B_{DL}$) and field-like ($B_{FL}$) effective fields as a function of Mo thickness in Mo(*y*)/CoGd(5) bilayers (Fig. 4a and Supplemental Material S11). The $B_{DL}$ increase monotonically with increasing Mo layer thickness (Fig. 4a), suggesting that the bulk OHE in the Mo layer plays a significant role in injecting strong OAM. Although $J_{OHE}$ from the Mo layer cannot directly interact with the magnetic moments, its conversion into SAM in the adjacent ferrimagnetic CoGd layer, enabled by strong SOC, enables high-efficiency deterministic switching, consistent with a recent report.[19,24,25,27]

Asymmetric spin transport across a single magnetic/nonmagnetic interface generates a finite *z*-polarized spin current, validated by PHE enhancement.[44-47] This yields a planar current density$\boldsymbol{J}_{\boldsymbol{PHE}}^{\boldsymbol{z}} = \Delta\sigma_{AMR}E(\boldsymbol{m}\cdot\hat{x})(\boldsymbol{m}\cdot\hat{z})$, generating a z-spin effective field.[34,44,45,46] Here $\Delta\sigma_{AMR}$ is the anisotropic magnetoresistance (AMR) conductivity and $\boldsymbol{E} = E\boldsymbol{x}$ is the electric field. Angle-dependent PHE resistivity $\rho_{PHE}(\theta)$ measurements provide direct signatures of symmetry breaking. A Ta(1)/CoGd(5) without a Mo layer exhibits only a weak PHE signal and no deterministic switching (Fig. 4b). In contrast, for Ta(0.5)/Mo(*y*)/CoGd(5), both the $\rho_{PHE}$ and anomalous Hall resistivity $\rho_{xy}(\varphi)$ increase substantially as *y* increases from 0 to 2 nm (Fig. 4b and Supplemental Material S12). For the Ta(0.5)/Mo(2)/CoGd(*t*) stack, PMA and deterministic switching vanish when $t < 3$ nm at 300 K, indicating a threshold thickness required to sustain the magnetic order necessary for PMA switching. For $t \geq 3$ nm, both $\rho_{PHE}$ and $\rho_{xy}(\varphi)$ increase only slightly (Supplemental Material 12). Thus, PHE enhancement predominantly originates from orbital contributions in the Mo layer.

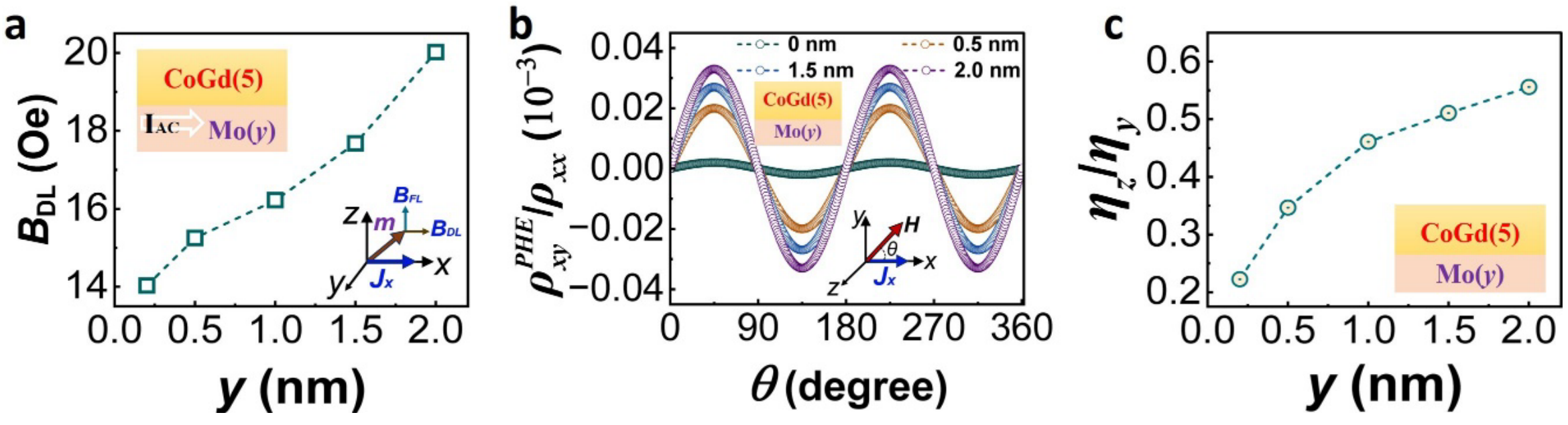

**FIG. 4.** (**a**) The estimated $B_{DL}$ effective field as a function of Mo layer thickness with a fixed switching current density of $J_x = 2.16 \times 10^6$ A cm$^{-2}$ and applying an in-plane $H_x$ field. (**b**) The PHE angular dependence on $\rho_{PHE}/\rho_{xx}$ ratio for Ta(1)/CoGd(5) (dark green color) and Ta(0.5)/Mo(*y*)/CoGd(5) samples in the *x*-*y*

plane, where $y$ is 0, 0.5, 1.5, and 2 nm at 300 K under a constant applied field of 1.4 T. (**c**) Ratio of out-of-plane to in-plane effective field components, $\eta_{\mathrm{z}}/\eta_{\mathrm{y}}$, for the Mo(y)/CoGd(5) samples with y varies from 0.2 to 2 nm.

The *z*-spin effective field ($\eta_z = \Delta H_z/J_x$) can be quantified by measuring the $R_H$ hysteresis loop shift at a positive and negative critical current densities $J_x^{\pm}$ (Supplemental Material S13). The AHE $R_H$-$H_z$ hysteresis loop shift is expressed as $\Delta H_z = H_s(J_x^+) - H_s(J_x^-)$, where $H_s(J_x^{\pm})$ is the ceneter of the hysteresis loop determined by the difference between positive and negative magnetization-reversal field values. Similarly, the *y*-spin effective field ($\eta_y = \Delta H_{DL}/J_c$) were measured by employing the standard harmonic Hall signal technique (Fig. 4a and Supplemental Material S11). The ratio$\eta_{\mathrm{z}}/\eta_{\mathrm{y}}$ between out-of-plane (*z*-spin) and in-plane (*y*-spin) components is plotted for Mo(y)/CoGd(5) samples with y varying from 0.2 to 2 nm (Fig. 4c). The ratio monotonically increases with increasing Mo layer thickness, revealing that the strong bulk OHE of the Mo layer largely enhances both *y*- and *z*-polarized spin.

Theoretical studies suggest that the intrinsic PHE can arise from orbital contributions in a weak-SOC system,[48] thereby generating a sufficient *z*-polarized spin that exceeds the intrinsic damping-like torque of the CoGd layer. The interfacial disorder in the OHE/ferrimagnetic system can further amplify the in-plane symmetry and, thereby, the PHE. In addition to the large $\eta_z$, the *y*-spin effective field in our Mo/CoGd bilayer is more than 4 times that of the CoGd/$CuO_x$ bilayer.[33] This indicates a much more efficient orbital-to-spin conversion in our system.

In conclusion, we have realized field-free deterministic magnetization switching driven by orbital-Hall-current-induced torques in a simple Mo/CoGd ferrimagnetic bilayer across a wide temperature range. This bilayer integrates the essential functions of an orbital-torque device: Mo generates a strong $\boldsymbol{J}_{OHE}$, whereas CoGd provides perpendicular ferrimagnetic switching and mediates orbital-to-spin conversion through its rare-earth transition-metal sublattices. The enhanced planar Hall response, current-induced anomalous Hall loop shifts, and harmonic Hall analysis reveal sizable in-plane and unconventional $z$-polarized damping-like torques. In particular, the $z$-polarized damping-like torque supplies the symmetry-breaking effective field required for deterministic switching without an external magnetic field. The combined thickness dependence shows that the switching efficiency is governed by the Mo orbital-current source, CoGd-mediated conversion, and Mo/CoGd interfacial symmetry breaking. As a result, the device reaches a critical switching current density down to $2.51 \times 10^6$ A cm$^{-2}$. These results demonstrate a compact route to orbital-current-induced switching in ferrimagnetic bilayers and advance the development of scalable, low-power orbitronic memory technologies.

## Methods

Film stacks consisting of Ta(0.5)/Mo($y$)/$Co_{71}Gd_{29}$($t$)/Ru(3)/Ta(2) (thickness in nm) were deposited onto a thermally oxidized Si/$SiO_2$ substrate using d.c. Magnetron sputtering at room temperature with a base pressure of less than $2.5 \times 10^{-8}$ Torr. Here, $y$ varied from 0.2 to 4 nm, and $t$ varied from 3 to 5 nm. In the body, we describe it as the Mo/CoGd bilayer. The Ta adhesion and Ru/Ta capping layer prevented oxidation. CoGd alloy films were obtained by co-deposition of Co and Gd targets at 35 W and 14 W, respectively, yielding $Co_{71}Gd_{29}$ alloy. The details of the alloy preparation method are described in the previous work.[9,30] All the sample thin films were patterned into 15 μm-wide Hall-bar devices using lithography and Ar ion etching.

The transport measurements were conducted using a Keithley 2602B and 2182 nanovoltmeters as a current source and a Hall voltage detector, respectively. A small current intensity of 50 μA was applied to measure the out-of-plane anomalous Hall voltages. The harmonic signals were acquired using SR830 DSP lock-in amplifier generator with a low frequency (13.333 Hz) and an alternating current ($I_{ac}$).

## Supplementary Material

See the supplementary material for additional information.

## Acknowledgments

K.W. and Z.A.B. gratefully acknowledge Dr. Dongwook Go and Prof. Yuriy Mokrosov for their insightful discussions and expert guidance. This research was supported by the National Key R&D Program of China (Grant Nos. 2022YFA1405100, 2024YFA1408503), the Beijing Natural Science Foundation for International Scientists Project (Funding No. IS23028), the NSFC (Grant Nos. 12241405, 12427805, 12104449, 52250418, 12274411, and 52501308), and the Natural Science Foundation of Shandong Province (No. ZR2025QC1106).

## Author Declarations

### A. Author Contributions

**Wang K:** conceived and supervised this work. **Bekele ZA:** grew the films, fabricated the devices, carried out the electrical transport measurements, and measured the magnetic properties. **Zhang L:** performed the theoretical analysis. **Bekele ZA**, **Cai. K**, **Zhang L**, and **Wang K:** analyzed the data and wrote the manuscript. All authors discussed the results and commented on the manuscript.

### B. Data Availability

The data supporting the findings of this study are available in the article and its supplementary material.

## Ethics Declarations

### Competing Interests

The authors have no conflicts to disclose.